# Depths of Copernican Craters on Lunar Maria and Highlands

E. A. Feoktistova[1] · S. I. Ipatov[2]

**Abstract**
We present a study on the relationship between the ratio of the depth of a crater to its diameter and the diameter for lunar craters both on the maria and on the highlands. We consider craters younger than 1.1 billion years in age, i.e. of Copernican period. The aim of this work is to improve our understanding of such relationships based on our new estimates of the craters's depth and diameter. Previous studies considered similar relationships for much older craters (up to 3.2 billion years). We calculated the depths of craters with diameters from 10 to 100 km based on the altitude profiles derived from data obtained by the Lunar Orbiter Laser Altimeter (LOLA) onboard the Lunar Reconnaissance Orbiter (LRO). The obtained ratios $h/D$ of the depths $h$ of the craters to their diameters $D$ can differ by up to a factor of two for craters with almost the same values of diameters. The linear and power approximations (regressions) of the dependence of $h/D$ on $D$ were made for simple and complex Copernican craters selected from the data from Mazrouei et al. (Science 363:253–255, 2019) and Losiak et al. (Lunar Impact Crater Database, 2015). For the separation of highland craters into two groups based only on their dependences of $h/D$ on $D$, at $D < 18$ km these are mostly simple craters, although some complex craters can have diameters $D \geq 16$ km. Depths of mare craters with $D \leq 14$ km are greater than $0.15D$. Following Pike's (Lunar Planet Sci XII:845–847, 1981) classification, we group mare craters of $D < 15$ km as simple craters. Mare craters with $15 < D < 18$ km fit both approximation curves for simple and complex craters. Depths of mare craters with $D > 18$ km are in a better agreement with the approximation curve of $h/D$ versus $D$ for complex craters than for simple craters. At the same diameter, mare craters are deeper than highland craters at a diameter smaller than 30–40 km. For greater diameters, highland craters are deeper. The values of $h/D$ for our approximation curves are mainly smaller than the values of the curve by Pike (in: Roddy, Pepin, Merrill (eds) Impact and explosion cratering: planetary and terrestrial implications, University of Arizona Press, Tucson, 1977) at $D < 15$ km. Only for mare craters at $D < 11$ km, our approximation curve is a little higher than the curve by Pike (1977). For our power approximations, the values of $h/D$ obtained for complex craters are greater than those obtained by Pike (1981) at $D > 53$ km for highland craters, and at $D < 80$ km for mare craters.

**Keywords** Copernican craters · Depths · Diameters · Lunar mare and highlands · LOLA altimeter

✉ E. A. Feoktistova
   hrulis@yandex.ru

Extended author information available on the last page of the article

## 1 Introduction

One of the important morphometric characteristics of impact craters is the ratio of the depth of a crater to its diameter. The morphometry of lunar craters has been the subject of several studies (e.g., Pike 1977, 1981; Kalynn et al. 2013). Pike (1977, 1981) considered the dependences of a crater depth $h$ on its diameter $D$ for "fresh" (belonging to Eratosthenian or Copernican periods) lunar craters. The dependences of $h$ on $D$ were obtained by Pike (1977) for a large database of lunar craters on the entire surface of the Moon. The difference between the values of the $h/D$ ratio for complex lunar craters located on the highlands and the maria was noted in Pike (1981): it was found that the value of $h/D$ for complex lunar craters on the highlands to be higher than for craters located on the maria, the large, dark, basaltic plains on the Moon, formed by ancient volcanic eruptions.

Impact craters are classified into two groups according to their morphology: simple and complex craters. Simple lunar craters are relatively small, with smooth bowl shapes and depths of about 1/5–1/7 of the crater diameter (Skelton et al. 2003). Complex craters may feature a central uplift, typically broad flat shallow crater floors, and terraced walls. Pike (1981) found that complex craters can be as small as 15 km in diameter across mare surfaces, whereas the equivalent smallest highland crater is on average 21 km across. The transitional (simple/complex) craters are characterized by relatively smooth walls and a floor partly or completely covered by debris slumped from the crater walls.

In Kalynn et al. (2013), 111 "fresh" craters with diameters from 15 to 167 km were chosen among craters with an age of less than 3.2 billion years (belonging to the Copernican and Eratosthenes periods). A crater is considered fresh if one or more of the following attributes are observed: (1) impact melt on the crater floor and ejecta facies, (2) a well-defined crater rim and/or (3) rays in the ejecta blanket. Of the 111 craters, 80 craters display central peaks characteristic of complex craters. Kalynn et al. (2013) used WAC images to classify each crater as either complex or transitional. Both types exhibit terraced walls, but complex (or central peak) craters also feature a clear central peak protruding through the melt sheet. A total of 31 craters were classified as transitional because they were without a clear central peak. Only 16 of the 57 craters used by Pike (1981) were also included in their data set. Kalynn et al. (2013) used Lunar Orbiter Laser Altimeter gridded data to determine the rim-to-floor crater depths, as well as the height of the central peak above the crater floor. Their results indicate that craters on highland terrain are, on average, deeper and have higher central peaks than craters on mare terrain. Kalynn et al. (2013) suggested that such differences resulted from differences in bulk physical properties of the terrain types, and differences in layering in the target terrain may also contribute to differences in final crater morphology. A similar conclusion was reached by Chandnani et al. (2019) and Osinsky et al. (2019). In Chandnani et al. (2019) well-preserved lunar craters with a diameter of 15—20 km were investigated. According to their findings, the deepest craters of this range of diameters are located near the highlands-mare boundaries. In Chandnani et al. (2019), the formation of deeper craters in such areas is associated with a possible higher soil porosity. The morphometric parameters of 28 well-preserved lunar craters with a diameter range from 15 to 42 km were studied by Osinsky et al. (2019). They found that the craters of the maria have a shallower depth than the craters of the highlands with the same diameters. This difference was also explained by the properties of the underlying crater surface.

The *h/D* ratio for complex craters in Kalynn et al. (2013) turned out to be higher than the similar ratio from Pike (1981). This difference was explained by the fact that older craters and, correspondingly, heavily modified ones were considered by Pike (1981).

The separation of lunar craters with a diameter $D \geq 3$ km into three types (simple, transitional, and complex craters) was proposed in Kruger et al. (2018). It was found that the transition from simple to complex craters on the maria occurs at diameters of ~14 km, and on the highlands at diameters of ~17 km. The transition from transitional morphology to complex on the highlands also occurs at large diameters of ~28 km. For the maria, this value is 24 km. In Kruger et al. (2018) as in Osinsky et al. (2019), these differences in the morphology of the craters of the lunar maria and the highlands were explained by the difference in the geological structure of the underlying surface.

Ipatov et al. (2020a, b) concluded that the number of craters per unit area on the maria is higher than that on the highlands for Copernican lunar craters, which ages were taken from Losiak et al. (2015) and Mazrouei et al. (2019). The number of Copernican craters with diameters $D > 15$ km per unit area is higher on the maria than on the highlands by a factor of 7 and 2 for craters selected from Losiak et al. (2015) and Mazrouei et al. (2019), respectively. For craters with diameters larger than 30 km, the number of craters per unit area on the maria is higher than on the highlands by a factor of 2 for craters from Losiak et al. (2015) and by a factor of 4 for craters from Mazrouei et al. (2019). An analysis of the dependence of *h/D* on *D* was not carried out by Losiak et al. (2015) and Mazrouei et al. (2019). The data from these papers were used mainly as references to the age of the craters.

The previous studies of the depths of craters by Pike (1977, 1981) and Kalynn et al. (2013) were made for craters with an age up to 3.2 billion years. We study the relationship between the ratio of a depth of a crater to its diameter with the diameter for craters with an age less than 1.1 billion years. The depths of the craters with diameters from 10 to 100 km were calculated in this study based on the altitude profiles derived from data obtained by the Lunar Orbiter Laser Altimeter onboard the Lunar Reconnaissance Orbiter.

## 2 Relationship of the Ratio of the Depth of a Crater to its Diameter with the Diameter of the Crater with an Age Less Than 3.2 Billion Years

Based on the analysis of data on craters, Pike (1977, 1981) and Kalynn et al. (2013) derived formulas for the relationship between a crater depth and its diameter. Using Apollo photogrammetric data, Pike (1977) analyzed Copernican and Eratosthenian craters (with an age younger than 3.2 Gyr). He concluded that for the distribution of 171 craters smaller than about 15 km across a crater depth *h* (in km) is described by the expression (on the right we present his formula also in another form which is used in Sects. 3 and 4):

$$h = 0.196 D^{1.01}, \ h/D = 0.196 D^{0.01} \ \text{at} \ D < 15 \ \text{km}. \tag{1}$$

According to Pike (1977), the depth/diameter distribution of the 33 craters over about 15 km in diameter follows the expression:

$$h = 1.044 D^{0.301}, \ h/D = 1.044 D^{-0.699} \ \text{at} \ D \geq 15 \ \text{km}. \tag{2}$$

Pike (1977) concluded that large mare craters do not appear to differ systematically from similar-size highland craters. Formulas (1) and (2) apply to the both types of lunar craters. In later studies, Pike (1981) remarked that fresh complex craters on the lunar highlands are deeper than those on mare surfaces. He concluded that within the diameter interval from 15 to 21 km usually all mare craters are complex and almost all highland craters are simple. It is seen from Fig. 1 from Pike (1981) that the main difference in $h/D$ for mare and complex craters is at $15 < D < 21$ km, because the ratio $h/D$ is greater for simple craters than for highland craters. In the figure the values of the ratio $h/D$ for highland craters to that for mare craters were mostly less than 1.1 at $D > 21$ km and could be close to 2 at $15 < D < 21$ km. Pike (1981) calculated the following relationships for complex craters:

$$h = 1.028D^{0.317}, \; h/D = 1.028D^{-0.683} \text{ for highland craters} \quad (3)$$

and

$$h = 0.819D^{0.341}, \; h/D = 0.819D^{-0.659} \text{ for mare craters.} \quad (4)$$

Like Pike (1981), Kalynn et al. (2013) considered craters of the Copernican and Eratosthenes periods of formation, but they selected those craters that appeared to be 'fresher' (their selection criteria are mentioned in the Introduction). It was shown in Kalynn et al. (2013) that the $h/D$ value for fresh complex craters (for the Copernican and Eratosthenes formation periods) located on the highlands of the Moon is higher than for craters located on the maria. Kalynn et al. (2013) suggested that the difference in $h/D$ for highland and mare craters is related to properties of the subsurface. This ratio also depends on the various properties of the impactors, such as a density, velocity and angle of incidence, that formed the craters (Melosh 1989; Collins et al. 2005; Werner and Ivanov 2015; Minton et al. 2015; Ipatov et al. 2020b). According to Kalynn et al. (2013), for 80 complex craters, the dependence of $h/D$ on the diameter of the crater can be expressed by the following formulas:

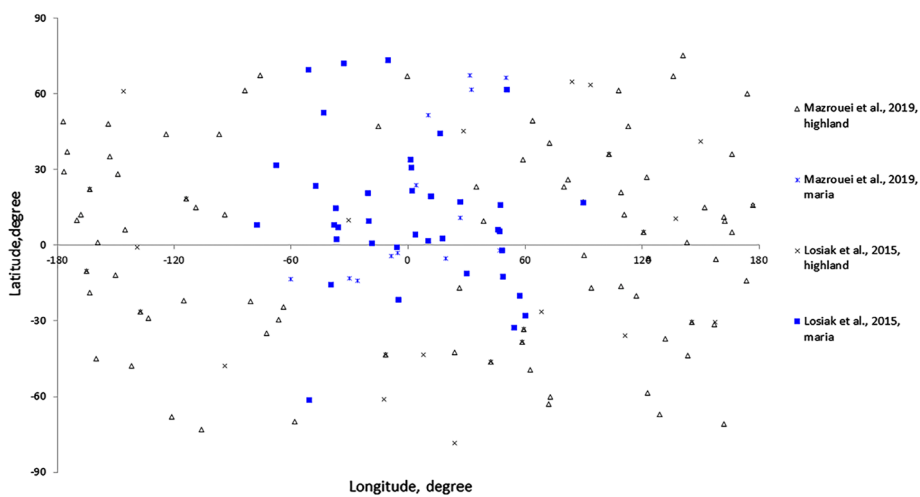

**Fig. 1** Coordinates of Copernican craters with a diameter $D \geq 10$ km taken from Losiak et al. (2015) and Mazrouei et al. (2019)

$$h = 1.558D^{0.254}, \ h/D = 1.558D^{-0.746} \ \text{for highland craters} \quad (5)$$

and

$$h = 0.870D^{0.352}, \ h/D = 0.870D^{-0.648} \ \text{for mare craters.} \quad (6)$$

Note that the formulas (5) and (6) were obtained for other ranges of diameters of craters than formulas (1)–(4) and the range of diameters considered by us. Among complex craters considered by Kalynn et al. (2013), there were only three craters with $D$ smaller than about 30 km (two mare craters with $D$ close to 30 km and one highland crater with $D \approx 25$ km), but the diameter of the largest crater was 167 km. All other considered craters with $D \leq 30$ km were considered by Kalynn et al. (2013) to be transitional, but not complex craters.

## 3 Relationship of the Ratio of the Depth of a Crater to its Diameter with the Diameter of the Crater for Craters with an Age of Less than 1.1 Billion Years

The ages of the craters were taken from Mazrouei et al. (2019) and Losiak et al. (2015). Pike (1977, 1981) and Kalynn et al. (2013) considered craters of the Copernican and Eratosthenes periods, i.e., their data could include older craters than the Copernican craters considered by Mazrouei et al. (2019) and Losiak et al. (2015). Distribution of Copernican craters with a diameter $D \geq 10$ km on the lunar surface is shown on Fig. 1. The data were taken from Mazrouei et al. (2019) and Losiak et al. (2015). Only 24 Copernican craters are presented in both papers. There are some discrepancies in the coordinates of the same craters given in these papers. For example, for the King crater the differences in longitude is up to 0.5 degrees. In this case, the coordinates of the craters presented in Mazrouei et al. (2019) were used in Fig. 1. More often craters presented in both papers are located closer to the equator. Most of them have a longitude between $-50°$ and $40°$. The distribution of craters of the Copernican age over the lunar surface according to data from Losiak et al. (2015) seems to be uneven: about 74% of such craters are in the lunar nearside. Copernican craters considered by Mazrouei et al. (2019) are distributed more evenly on the lunar surface: 48.6% and 51.4% are in the nearside and farside of the Moon, respectively.

To estimate the depths $h$ and diameters $D$ of the Copernican craters taken from Losiak et al. (2015) and Mazrouei et al. (2019), we constructed altitude profiles for each crater using the heights of the points of the lunar surface obtained by the LOLA altimeter (https://ode.rsl.wustl.edu/moon/). Based on these profiles, we calculated the crater diameter as the average value of the diameter in the directions of latitude and longitude. Such diameter in a direction was determined as the distance between points on the crest of the crater ridge (Pike 1977). A crater depth was determined as the distance between the rim of the crater and the crater floor. We used latitudinal and longitudinal altitude profiles of each crater to obtain the average values of the height of the rim crater above the surrounding surface and the average values of the level of the crater bottom. The values of the height were averaged over four maximum values of the height of the rim calculated based on these profiles. The profiles were also used for estimation of the average value of the level of the crater bottom. Latitudinal (a) and longitudinal (b) altitude profiles of the Proclus crater are shown in Fig. 2.

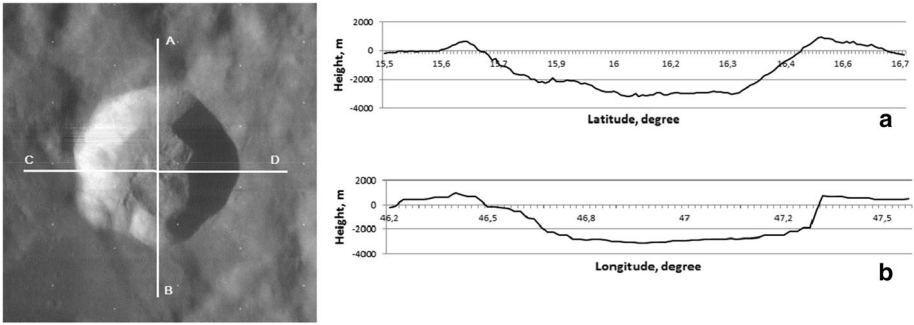

**Fig. 2** Latitudinal AB (**a**) and longitudinal CD (**b**) altitude profiles of the crater Proclus (Lunar Orbiter Photo Gallery, Mission 4)

We considered the Copernican craters separately for the maria and the highland from Losiak et al. (2015). The database of lunar craters from Losiak et al. (2015) contains 66 craters with a diameter greater than 10 km and with an age of less than 1.1 billion years, 38 of which are located on the nearside maria and 28 are on the highlands. As in (Ipatov et al. 2020b), the ratio of the area of the considered region of the maria to the area of the whole surface of the Moon is 0.155. Information on the age of the craters in the catalog (Losiak et al. 2015) is based on data from Wilhelms (1987) and Wilhelms and Byrne (2009).

Mazrouei et al. (2019) used an analysis of the thermophysical characteristics of lunar impact ejecta as measured with the Diviner thermal radiometer onboard the LRO probe to estimate the ages of lunar craters with a diameter $D > 10$ km and younger than 1 billion years. They concluded that formation of large lunar craters excavates numerous $\geq 1$ m ejecta debris onto the Moon's surface. These recently exposed rocks have high thermal inertia and remain warm during the lunar night relative to the surrounding lunar soils (called regolith), which have low thermal inertia. Young craters were found to have high rock abundance in their ejecta, whereas rock abundance decreases with increasing crater age, eventually becoming indistinguishable from the background for craters older than ~1 Gyr. Mazrouei et al. (2019) identified 111 Copernican craters on the Moon with $10 \leq D < 100$ km between 80°N and 80°S, with ejecta blankets that have rock abundance values high enough to distinguish them from the background regolith. However, the resulting sample of craters included only 24 craters, whose age was defined as less than 1.1 billion years according to earlier studies (Wilhelms 1987; Wilhelms and Byrne 2009).

The distribution of Copernican craters considered by Losiak et al. (2015) and Mazrouei et al. (2019) over diameter and type of target surface (the maria and the highlands) is shown in Fig. 3. The largest number of Copernican craters presented in this figure was based on data from Mazrouei et al. (2019), have a diameter of $10 \leq D < 20$ km and are located on the highlands. According to data from Losiak et al. (2015), the number of lunar craters of this age on the maria is greater than or equal to the number of craters on the highlands for all intervals of diameters, except for 80–90 km. Ipatov et al. (2020b) noted that a comparison of the distribution of craters of Copernican age with $D < 30$ km from Losiak et al. (2015) with the curve from Neukum et al. (2001) showed that these data are possibly incomplete.

Figure 4 shows the dependence of *h/D* on the diameter of the crater. For craters with an age of less than 3.2 billion years, the dependences are given by formulas (2)-(6). Formula (2) was obtained for the entire surface of the Moon, formulas (3) and (5) are for the highlands, and formulas (4) and (6) are for the maria. For Copernican craters selected from Losiak et al. (2015) and Mazrouei et al. (2019), we also present the $h/D$ dependences for Copernican craters located on the highlands and on the maria. The depths and diameters of the craters taken from Mazrouei et al. (2019) and Losiak et al. (2015) were calculated based on the LOLA LRO elevation profiles. We consider plots of $h/D$ versus $D$ because we believe that such dependences better characterize our studies of the morphology of craters than the plots of $h$ versus $D$.

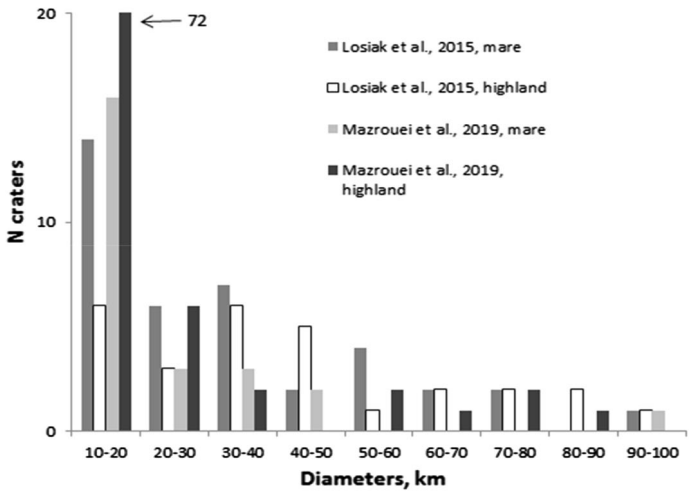

**Fig. 3** The histogram of the distribution of the Copernican craters, which age was taken from Losiak et al. (2015) and Mazrouei et al. (2019), over a diameter and type of underlying surface (mare or highland). Of the four close columns, two left columns are data from Losiak et al. (2015). There are 72 highland craters with $10 \leq D < 20$ km from Mazrouei et al. (2019)

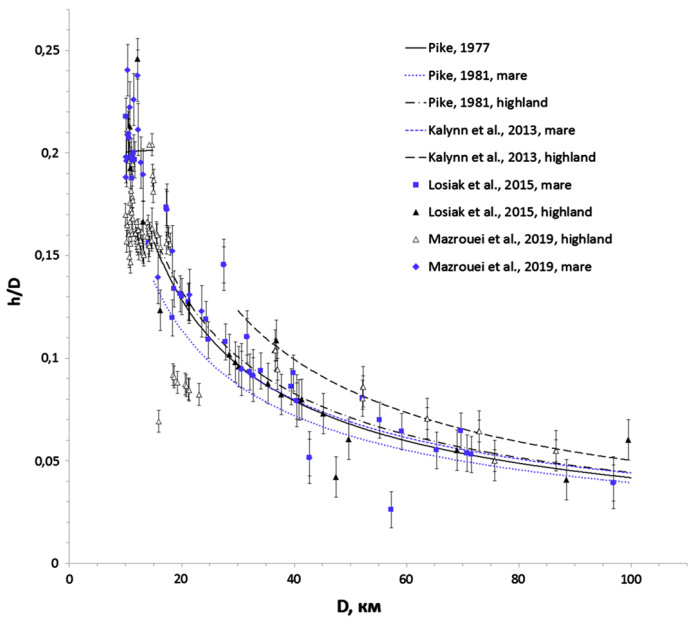

**Fig. 4** Relationship of the $h/D$ ratio of a crater depth to diameter with the diameter. For craters with an age of less than 3.2 billion years, the plots are presented according to Pike (1977) (formula (2)) for the entire surface of the Moon, and according to Pike (1981) (formulas (3) - (4)) and Kalynn et al. (2013) (formulas (5)–(6)) for the region of the highlands and the maria. For craters with an age of less than 1.1 billion years selected from Losiak et al. (2015) and Mazrouei et al. (2019), the plots are for the region of the highlands and the maria. Depths and diameters of such craters were calculated by us based on the LOLA LRO S–N and E–W profiles of the craters. Error bars show standard error of the mean for craters from Mazrouei et al. (2019) and Losiak et al. (2015)

The ratio $h/D$ of the depth of the crater to its diameter mostly decreases with the increase of a diameter $D$ of the crater for all dependencies (see Fig. 4). For the curves from Kalynn et al. (2013) and Pike (1981), the $h/D$ values for the highlands are higher than for the maria, and the curves from Kalynn et al. (2013) are higher than the corresponding curves from Pike (1981). The difference between the $h/D$ values for the highlands and the maria for the curves from Kalynn et al. (2013) is several times greater than for the curves from Pike (1981). As noted in the previous section, approximation curves from Kalynn et al. (2013) are for craters that, on average, have larger diameters and younger ages than those considered by Pike (1981). Kalynn et al. (2013) considered mainly craters with $D > 30$ km. As it is seen from Fig. 4, the dependence of $h/D$ on $D$ is steeper at $D < 30$ km than at $D > 30$ km.

Crater depths calculated based on LOLA LRO altitude profiles can vary significantly for craters with similar diameters. For example, for craters on the maria taken from Mazrouei et al. (2019), the $h/D$ ratio is 0.145, 0.094 and 0.110 at $D$ equal to 27.5, 30.7 and 31.7 km, respectively. For craters on the highland taken from Mazrouei et al. (2019), the $h/D$ ratio is equal to 0.092 and 0.156 at $D = 18.5$ and $D = 18.0$ km, i.e. $h/D$ can vary by a factor of 1.7 for almost the same $D$. A similar scatter of $h/D$ values was obtained for some craters taken from Losiak et al. (2015). For example, for craters on the maria taken from Losiak et al. (2015), the average values of $h/D$, each for two craters, were 0.173 and 0.127 for intervals of diameters of 17.3–17.4 and 18.4–18.7 km, respectively, that is, they differed by a factor of 1.36. For three mare craters with diameters equal to 55.2, 57.3 and 59.2 km, the $h/D$ ratio was 0.070, 0.026, and 0.064, respectively, and the values of $h/D$ differ by a factor of 2.7. The same large scatter in $h/D$ values was obtained for craters on the highlands taken from Losiak et al. (2015). For such craters, the values of $h/D$ were 0.042 and 0.073 at $D$ equal to 47.4 and 45.2 km. That is, the $h/D$ values at close values of $D$ could differ by about a factor of 1.7. The wide scatter of crater depths at similar diameters calculated with the use of LOLA LRO altitude profiles is due to the fact that the difference in crater depths at almost the same diameter in some cases exceeds 1 km. The difference in the depth of craters with similar diameters can be explained by the difference in the parameters of the impactors that formed the given crater: the ratio of the density of the impactor to the density of the target, the velocity of the impactor, and the angle of incidence. After the formation of the crater, it can be subjected to other influences: the fall of new projectiles or geological processes. This can lead to the appearance of lava at the bottom or the collapse of the walls of the crater and lead to a change in its depth.

For estimates of crater depths, Losiak et al. (2015) used previous data from Arthur (1974), Pike (1976), and Wood (1973). The difference between the depths of the craters on the maria and on the highlands for these data was negligible. If we consider these data, we obtain relatively smooth curves of $h/D$ versus $D$ for most values of $D$ (they are not shown in Fig. 4). In this case, for $18 < D < 98$ km, the curves of $h/D$ versus $D$ are close to two parallel segments of curves with a small jump in the $h/D$ values at $D \approx 58$ km. For these curves, a sharp jump in $h/D$ values was obtained at $D = 18$ km: $h/D \geq 0.16$ at $D \leq 17.6$ km and $h/D \leq 0.09$ at $D \geq 18.8$ km. The jump might correspond to the transition from simple to complex craters. When applying depth estimates for craters from (Arthur 1974; Pike 1976;

Wood [1973]) to the data from Losiak et al. ([2015]), the curve of $h/D$ versus $D$ lies below the curves based on formulas from Kalynn et al. ([2013]) and Pike ([1977], [1981]).

In Fig. [4] the ratio $h/D$ for craters from Mazrouei et al. ([2019]) and Losiak et al. ([2015]) is either higher or lower than the curves of $h/D$ versus $D$ from Pike ([1977]) and Pike ([1981]). These values mainly lie below the curve from Kalynn et al. ([2013]) for highland craters. Therefore, our estimates are closer to those by Pike ([1981]) than Kalynn et al. ([2013]). As seen from Fig. [4], for highland craters from Mazrouei et al. ([2019]) at $D > 35$ km, $h/D$ ratios are usually greater than those for craters on the maria as from Mazrouei et al. ([2019]) and Losiak et al. ([2015]).

The average values of $h/D$ obtained for several close values of $D$ are shown in Table [1]. The craters considered in Table [1] were taken from Mazrouei et al. ([2019]) and Losiak et al. ([2015]). In Table [1]a the values of diameters of craters were based on the above papers, and our estimates of the diameters are used in Table [1]b. Our estimates of depths of craters were used for both Table [1]a, b. In Table [1]a the sizes of considered $D$ intervals were chosen in such a way that the values of $h/D$ were about the same for such interval. This is a histogram binning with variable sizes of $D$ intervals. In Table [1]b $D$ interval is the same (equaled to 2.5, 5, 10 or 20 km) for some values of $D$, but typically is greater for greater $D$. Sizes of two short intervals of $D$ (17.6–18.4 and 18.5–20 km) were chosen so that the ratio $h/D$ was about the same for $D$ inside each interval. The number of considered craters for each $D$ interval is given in parentheses in Table [1]. For craters from Losiak et al. ([2015]) with $D > 40$ km, the averaged values of $h/D$ in Table [1]a for craters on the highlands are no less than those for craters on the maria (are greater for some values of $D$ and about the same for other values of $D$). In Mazrouei et al. ([2019]), there are not enough craters with diameters $D > 40$ km needed to compare the number of craters on the highlands and the maria. For craters with diameters from 10 to 30 km, in most cases, on the contrary, the depths of craters on the maria are no less than the depths of craters on the highlands. In Table [1]b at $D > 15$ km the number of craters on the highlands does not differ much from the number of craters on the maria, though the area of the highlands is greater by a factor of 5.45 than the area for the maria. The mean values of $h/D$ for $D$ intervals in Table [1]b are mostly lower for the highlands than for the maria at $D < 30$ km and are greater at $D > 30$ km. For $D$ interval from 30 to 40 km, the difference between mean values of $h/D$ is only 1%.

For dependences (3)–(4) taken from Pike ([1981]), the difference between the values of $h/D$ for craters on the maria and highlands is relatively small, and it can be smaller than the maximum difference in the values of this ratio for close values of $D$ if we calculate crater depths using LOLA LRO altitudinal profiles for craters from Mazrouei et al. ([2019]) and Losiak et al. ([2015]).

## 4 Comparison of the Formulas for the Relationship of the Ratio of the Depth of a Crater to its Diameter with the Diameter

### 4.1 Construction of Formulas and Plots for the Relationship of the Ratio of the Depth of a Crater to its Diameter with the Diameter

Analyzing the data for Copernican craters discussed in the previous section, we obtained the formulas for an approximation of the relationship of the ratio $h/D$ of the depth of a crater to its diameter with the diameter for craters on the highlands and the maria. For these approximations (regression) we used the software from https://plane

**Table 1** The mean values of the ratio $h/D$ of a depth of a crater to its diameter over intervals of the diameter $D$ for Copernican craters on the lunar maria and the highlands

(a)

| Interval of $D$, km | Losiak et al., highland | Losiak et al., mare | Interval of $D$, km | Mazrouei et al., highland | Mazrouei et al., mare |
|---|---|---|---|---|---|
| 10.0–12.2 | **0.216** (3) > | 0.207 (7) | 10–12 | 0.164 (37) < | **0.205** (9) |
| 13.5–16 | 0.149 (3) < | **0.159** (1) | 13–17 | 0.159 (29) < | **0.187** (5) |
| 17.0–17.6 | – | 0.156 (3) | | | |
| 18.8–19.9 | – | 0.091 (3) | 18–19 | 0.122 (6) < | **0.156** (2) |
| 21–29.4 | 0.106 (4) < | **0.118** (6) | 21–28 | 0.083 (6) < | **0.132** (3) |
| 30.8–39.8 | 0.093 (6) < | **0.096** (6) | 32–38 | 0.10 (2) = | 0.096 (3) |
| 40–48.3 | **0.072** (4) > | 0.067 (1) | 40–43 | – | 0.066 (2) |
| 50–59.1 | 0.060 (1) = | 0.061 (4) | 52–53 | 0.083 (2) | – |
| 61.8–64 | **0.073** (1) > | 0.057 (1) | 64 | 0.070 (1) | – |
| 70.2–76.2 | 0.057 (3) = | 0.057 (3) | 71–76 | 0.058 (2) | – |
| 85.3–86.2 | 0.047 (2) | – | 86 | 0.055 (1) | – |
| 96.1–98.2 | **0.061** (1) > | 0.040 (1) | 97 | – | 0.039 (1) |

(b)

| Interval of $D$, km | Losiak et al. + Mazrouei et al., highland | Losiak et al. + Mazrouei et al., mare |
|---|---|---|
| 10.0–12.5 | 0.16346 (37) < | **0.2036** (21) |
| 12.6–15 | 0.1657 (23) < | **0.1742** (4) |
| 15.1–17.5 | 0.1374 (6) < | **0.1646** (4) |
| 17.6–18.4 | **0.1576** (3) ≥ | 0.1522 (1) |
| 18.5–20 | 0.0903 (3) < | **0.1329** (3) |
| 20.1–25 | 0.0847 (6) < | **0.1231** (6) |
| 25.1–30 | 0.0999 (2) < | **0.1266** (2) |
| 30.1–40 | **0.0955** (6) ≥ | 0.0946 (7) |
| 40.1–50 | **0.0670** (5) > | 0.0654 (2) |
| 50.1–60 | **0.0832** (2) > | 0.601 (4) |

**Table 1** (continued)

| Interval of $D$, km | Losiak et al. + Mazrouei et al., highland | Losiak et al. + Mazrouei et al., mare |
|---|---|---|
| 60.1–70  | **0.0628** (2) > | 0.0598 (2) |
| 70.1–80  | **0.0573** (2) > | 0.0534 (2) |
| 80.1–100 | **0.0519** (3) > | 0.0392 (1) |

The number of craters in the considered interval is given in parentheses. The craters were selected based on the data from Losiak et al. (2015) and Mazrouei et al. (2019). The larger values of $h/D$ for the highland or the mare are in bold numbers. In Table 1a the diameters of considered craters were taken from the above papers, and in Table 1b the diameters were calculated by us

**Table 2** Formulas for the approximation of the dependences of the ratio $y = h/D$ of the depth of a crater to its diameter on the diameter $x = D$ for craters on the highlands

| Paper | Type of surface | Range of diameters | Range of $h/D$ | Number of craters | Linear function | Error, % | Power function | Error, % |
|---|---|---|---|---|---|---|---|---|
| Losiak | Highland | 10.6–98.2 | 0.042–0.246 | 28 | $y = -0.0016x + 0.1711$ | 26.4 | $y = 0.975x^{-0.666}$ | 12.22 |
| Losiak | Highland | 10.6–15 | 0.16–0.246 | 5 | $y = -0.0122x + 0.3470$ | 9.21 | $y = 1.509x^{-0.819}$ | 9.06 |
| Losiak | Highland | 10.6–16 | 0.125–0.246 | 6 | $y = -0.0156x + 0.3872$ | 9.63 | $y = 3.587x^{-1.171}$ | 10.18 |
| Losiak | Highland | 16–98.2 | 0.125–0.246 | 12 | $y = -0.00090x + 0.1248$ | 15.51 | $y = 0.724x^{-0.591}$ | 12.55 |
| Losiak | Highland | 22.1–98.2 | 0.122–0.246 | 11 | $y = -0.00085x + 0.1216$ | 15.71 | $y = 0.835x^{-0.627}$ | 12.40 |
| Mazrouei | Highland | 10–86 | 0.050–0.2 | 86 | $y = -0.00188x + 0.1815$ | 15.04 | $y = 0.666x^{-0.567}$ | 12.69 |
| Mazrouei | Highland | 10–18 | 0.069–0.2 | 69 | $y = -0.00238x + 0.1946$ | 7.60 | $y = 0.297x^{-0.236}$ | 7.61 |
| Mazrouei | Highland | 10–18 | 0.15–0.2 | 68 | $y = -0.00138x + 0.1831$ | 5.95 | $y = 0.221x^{-0.115}$ | 5.88 |
| Mazrouei | Highland | 19–86 | 0.050–0.103 | 17 | $y = -0.00046x + 0.0989$ | 7.83 | $y = 0.185x^{-0.239}$ | 9.47 |
| Mazrouei | Highland | 19–24 | 0.079–0.089 | 9 | $y = -0.00201x + 0.1277$ | 0.52 | $y = 0.391x^{-0.500}$ | 0.67 |
| Mazrouei | Highland | 36–86 | 0.050–0.103 | 8 | $y = -0.00101x + 0.1362$ | 5.99 | $y = 1.586x^{-0.759}$ | 6.15 |
| Losiak+Mazrouei | Highland | 10–98.2 | 0.042–0.246- | 114 | $y = -0.00181x + 0.1796$ | 18.36 | $y = 0.700x^{-0.582}$ | 12.90 |
| Losiak+Mazrouei | Highland | 10–18 | 0.125–0.2 | 75 | $y = -0.00327x + 0.2077$ | 8.58 | $y = 0.353x^{-0.302}$ | 8.46 |
| Losiak+Mzarouei | Highland | 19–98.2 | 0.042–0.122 | 39 | $y = -0.00064x + 0.1087$ | 13.77 | $y = 0.310x^{-0.376}$ | 15.93 |
| *Losiak+Mazrouei | Highland | 10–98 | 0.042–0.2 | 102 | $y = -0.00185x + 0.1796$ | 18.82 | $y = 0.718x^{-0.594}$ | 13.00 |
| *Losiak+Mazrouei | Highland | 10–15 | 0.153–0.2 | 61 | $y = -0.00030x + 0.1716$ | 7.32 | $y = 0.184x^{-0.039}$ | 7.10 |
| *Losiak+Mazrouei | Highland | 10–18 | 0.069–0.2 | 71 | $y = -0.00289x + 0.2019$ | 8.40 | $y = 0.329x^{-0.275}$ | 8.33 |
| *Losiak+Mazrouei | Highland | 10–18 | 0.065–0.2 | 69 | $y = -0.00156x + 0.1865$ | 6.55 | $y = 0.228x^{-0.125}$ | 6.38 |
| *Losiak+Mazrouei | Highland | 16–98 | 0.042–0.159 | 41 | $y = -0.00105x + 0.1335$ | 22.09 | $y = 0.491x^{-0.497}$ | 19.89 |
| *Losiak+Mazrouei | Highland | 19–98.2 | 0.042–0.108 | 31 | $y = -0.00058x + 0.1038$ | 13.09 | $y = 0.271x^{-0.347}$ | 15.10 |

The Copernican craters and their diameters were considered based on the data from Mazrouei et al. (2019) and Losiak et al. (2015) or from both papers. The lines corresponding to data from both papers are denoted as "Losiak+Mazrouei". For lines marked by * and located at the bottom of the table, we consider one point $(D, h/D)$ for the crater which was considered in both papers. The value of $D$ for such points was taken from Mazrouei et al. (2019). For lines marked by "Losiak+Mazrouei" (without *) for the same crater, we took two points $(D, h/D)$, based on the values of diameters of craters presented in both papers. For estimates of the depths $h$ of the craters, we used altitude profiles based on altimeter data of LOLA onboard the probe LRO (Lunar Reconnaissance Orbiter). The linear and power functions were used for the approximation of the values $h/D$ for several ranges of $D$. The average approximation errors for such approximations are presented

**Table 3** Formulas for the approximation of the dependences of the ratio $y = h/D$ of the depth of a crater to its diameter on the diameter $x = D$ for craters on the highlands

| Paper | Type of surface | Range of diameters | Range of $h/D$ | Number of craters | Linear function | Error, % | Power function | Error, % |
|---|---|---|---|---|---|---|---|---|
| Losiak + Mazrouei | Highland | 10–99.6 | 0.041–0.212- | 100 | $y = -0.001845x + 0.1797$ | 19.15 | $y = 0.726x^{-0.597}$ | 13.11 |
| Losiak + Mazrouei | Highland | 10–15 | 0.147–0.212 | 60 | $y = -0.000609x + 0.1759$ | 7.52 | $y = 0.193x^{-0.039}$ | 7.25 |
| Losiak + Mazrouei | Highland | 15.5–99.6 | 0.042–0.161 | 40 | $y = -0.001023x + 0.1319$ | 21.69 | $y = 0.478x^{-0.490}$ | 19.69 |
| -Losiak + Mazrouei | Highland | 10–18.0 | 0.123–0.212 | 68 | $y = -0.00198x + 0.1921$ | 6.96 | $y = 0.249x^{-0.160}$ | 6.72 |
| **= Losiak + Mazrouei** | **Highland** | **10–18.0** | **0.069–0.212** | **69** | $\mathbf{y = -0.00298x + 0.2037}$ | **8.73** | $\mathbf{y = 0.334x^{-0.281}}$ | **8.59** |
| Losiak + Mazrouei | Highland | 10–18.7 | 0.069–0.212 | 71 | $y = -0.00484x + 0.2264$ | 10.44 | $y = 0.539x^{-0.473}$ | 10.53 |
| Losiak + Mazrouei | Highland | 19.3–99.6 | 0.041–0.109 | 29 | $y = -0.00059x + 0.1046$ | 13.45 | $y = 0.298x^{-0.371}$ | 15.40 |
| **= Losiak + Mazrouei** | **Highland** | **18.5–99.6** | **0.041–0.109** | **31** | $\mathbf{y = -0.000580x + 0.1042}$ | **12.65** | $\mathbf{y = 0.277x^{-0.353}}$ | **14.77** |

Designations are the same as in Table 2. In contrast to Table 2, the values of diameters of considered craters are based on our calculations. The lines in the table for which there are plots in Figs. 5 and 6 are marked by = and bold text. For the line "-Losiak + Mazrouei" at 68 craters, the crater with $D = 15.9$ km and $h/D = 0.069$ was excluded compared to the below line with 69 craters

Table 4 Formulas for the approximation of the dependences of the ratio $y = h/D$ of the depth of a crater to its diameter on the diameter $x = D$ for craters on the maria

| Paper | Type of surface | Range of diameters | Range of $h/D$ | Number of craters | Linear function | Error, % | Power function | Error, % |
|---|---|---|---|---|---|---|---|---|
| Losiak | Mare | 10–96.1 | 0.026–0.22 | 37 | $y = -0.00218x + 0.1918$ | 26.38 | $y = 1.203x^{-0.647}$ | 10.89 |
| Losiak | Mare | 10–11.6 | 0.193–0.22 | 8 | $y = -0.0113x + 0.3248$ | 2.31 | $y = 0.842x^{-0.600}$ | 2.22 |
| Losiak | Mare | 10–13.8 | 0.159–0.22 | 9 | $y = -0.0139x + 0.3528$ | 2.21 | $y = 1.705x^{-0.895}$ | 2.31 |
| Losiak | Mare | 10–17.5 | 0.159–0.22 | 11 | $y = -0.00511x + 0.2549$ | 3.68 | $y = 0.502x^{-0.387}$ | 3.54 |
| Losiak | Mare | 13.8–96.1 | 0.026–0.173 | 30 | $y = -0.00161x + 0.1608$ | 20.15 | $y = 1.360x^{-0.769}$ | 13.42 |
| Losiak | Mare | 17.3–96.1 | 0.026–0.173 | 29 | $y = -0.00156x + 0.1582$ | 20.00 | $y = 1.449x^{-0.647}$ | 13.61 |
| Losiak | Mare | 17.6–96.1 | 0.026–0.149 | 27 | $y = -0.00138x + 0.1477$ | 18.02 | $y = 1.291x^{-0.756}$ | 13.27 |
| Mazrouei | Mare | 10–97 | 0.039–0.227 | 25 | $y = -0.00252x + 0.2137$ | 27.02 | $y = 1.318x^{-0.764}$ | 10.56 |
| Mazrouei | Mare | 10–13 | 0.173–0.227 | 13 | $y = -0.00134x + 0.2195$ | 6.74 | $y = 0.245x^{-0.076}$ | 6.69 |
| Mazrouei | Mare | 17–97 | 0.039–0.158 | 12 | $y = -0.00145x + 0.1567$ | 23.10 | $y = 1.729x^{-0.841}$ | 11.54 |
| Mazrouei | Mare | 21–97 | 0.039–0.143 | 9 | $y = -0.00120x + 0.1416$ | 23.41 | $y = 2.117x^{-0.895}$ | 12.63 |
| Losiak+Mazrouei | Mare | 10–97 | 0.026–0.227 | 63 | $y = -0.00237x + 0.2029$ | 27.50 | $y = 1.253x^{-0.748}$ | 10.99 |
| Losiak+Mazrouei | Mare | 10–13.8 | 0.159–0.227 | 22 | $y = -0.00531x + 0.2613$ | 5.69 | $y = 0.442x^{-0.327}$ | 5.61 |
| Losiak+Mazrouei | Mare | 17–97 | 0.026–0.173 | 41 | $y = -0.00153x + 0.1579$ | 21.11 | $y = 1.512x^{-0.799}$ | 13.08 |
| *Losiak+Mazrouei | Mare | 10–97 | 0.026–0.227 | 51 | $y = -0.00247x + 0.2065$ | 25.26 | $y = 1.238x^{-0.744}$ | 10.67 |
| *Losiak+Mazrouei | Mare | 10–97 | 0.039–0.227 | 50 | $y = -0.00241x + 0.2057$ | 22.41 | $y = 1.118x^{-0.705}$ | 7.99 |
| *Losiak+Mazrouei | Mare | 10–13.8 | 0.159–0.227 | 19 | $y = -0.00523x + 0.2603$ | 6.47 | $y = 0.437x^{-0.322}$ | 6.42 |
| *Losiak+Mazrouei | Mare | 10–19 | 0.133–0.227 | 24 | $y = -0.00730x + 0.2836$ | 7.50 | $y = 0.820x^{-0.582}$ | 7.68 |
| *Losiak+Mazrouei | Mare | 19.5–97 | 0.026–0.143 | 27 | $y = -0.00138x + 0.1476$ | 18.57 | $y = 1.494x^{-0.794}$ | 12.92 |
| *Losiak+Mazrouei | Mare | 19.5–97 | 0.039–0.143 | 26 | $y = -0.00131x + 0.1465$ | 12.54 | $y = 1.203x^{-0.726}$ | 7.85 |
| *Losiak+Mazrouei | Mare | 17–97 | 0.026–0.171 | 32 | $y = -0.00159x + 0.1598$ | 19.01 | $y = 1.463x^{-0.789}$ | 12.55 |
| *Losiak+Mazrouei | Mare | 17–97 | 0.039–0.171 | 31 | $y = -0.00152x + 0.1586$ | 14.09 | $y = 1.213x^{-0.789}$ | 8.31 |

Designations are the same as in Table 2

**Table 5** Formulas for the approximation of the dependences of the ratio $y = h/D$ of the depth of a crater to its diameter $x = D$ for craters on the maria

| Paper | Type of surface | Range of diameters | Range of $h/D$ | Number of craters | Linear function | Error, % | Power function | Error, % |
|---|---|---|---|---|---|---|---|---|
| Losiak + Mazrouei | Mare | 10.1–96.9 | 0.026–0.240 | 59 | $y = -0.002547x + 0.2106$ | 23.75 | $y = 1.218x^{-0.739}$ | 10.11 |
| =**Losiak + Mazrouei** | **Mare** | **10.1–15.8** | **0.139–0.240** | **26** | $y = -0.01056x + 0.3198$ | **6.49** | $y = 1.124x^{-0.715}$ | **6.74** |
| Losiak + Mazrouei | Mare | 10.1–18.7 | 0.134–0.240 | 32 | $y = -0.007469x + 0.2836$ | 7.22 | $y = 0.828x^{-0.587}$ | 7.29 |
| Losiak + Mazrouei | Mare | 19.8–96.9 | 0.026–0.145 | 27 | $y = -0.00138x + 0.1475$ | 18.29 | $y = 1.494x^{-0.794}$ | 12.90 |
| -Losiak + Mazrouei | Mare | 19.8–96.9 | 0.039–0.143 | 26 | $y = -0.00131x + 0.1464$ | 12.36 | $y = 1.205x^{-0.726}$ | 7.86 |
| =**Losiak + Mazrouei** | **Mare** | **17.3–96.9** | **0.026–0.173** | **33** | $y = -0.001679x + 0.1650$ | **19.70** | $y = 1.589x^{-0.810}$ | **12.27** |
| -Losiak + Mazrouei | Mare | 17.3–96.9 | 0.039–0.173 | 32 | $y = -0.001612x + 0.1638$ | 15.04 | $y = 1.325x^{-0.751}$ | 8.28 |

Designations are the same as in Table 2. In contrast to Table 4, the values of diameters of considered craters are based on our calculations. The lines in the table for which there are plots in Figs. 5 and 6 are marked by = and bold text. For lines "Losiak + Mazrouei" at the range of diameters (19.8–96.9) and 26 craters, and at the range of diameters (17.3–96.9) and 32 craters, the crater with $D = 57.3$ km and $h/D = 0.026$ was excluded

tcalc.com/5992/. The results of such approximation are presented in Tables 2, 3, 4 and 5. We consider craters on the highlands in Tables 2 and 3 and craters on the maria in Tables 4 and 5. The Copernican craters were selected based on the data from Mazrouei et al. (2019) and Losiak et al. (2015). For estimates of the depths $h$ of these craters, we used altitude profiles derived from altimeter data of LOLA (the Lunar Orbiter Laser Altimeter) onboard LRO (Lunar Reconnaissance Orbiter). In Tables 2 and 4, the values of diameters of craters are based on the above papers. In Tables 3 and 5, the values of the diameters were calculated by us.

In different lines of Tables 2, 3, 4 and 5, we consider data for craters with a different range of diameters. The range of the ratio $h/D$ for craters with such diameters and the number $N_{set}$ of such craters are also presented in the tables. For approximations for each set of craters, we used both a linear and a power function. A linear regression equation was considered in the form

$$\hat{y} = b \cdot x + c, \tag{7}$$

where

$$\text{where} \quad b = \left[\sum x_i \sum y_i - n \sum x_i y_i\right] / \left[\left(\sum x_i\right)^2 - n \sum x_i^2\right],$$
$$c = \left[\sum x_i \sum x_i y_i - \sum x_i^2 \sum y_i\right] / \left[\left(\sum x_i\right)^2 - n \sum x_i^2\right]. \tag{8}$$

A power regression equation was considered in the form

$$\hat{y} = k \cdot x^{\alpha}, \tag{9}$$

$$\text{where} \quad k = \exp\left(n^{-1}\sum \ln y_i - n^{-1}\alpha \sum \ln x_i\right),$$
$$\alpha = \left[n \sum (\ln x_i \ln y_i) - \sum \ln x_i \sum \ln y_i\right] / \left[n \sum \ln^2 x_i - \left(\sum (\ln x_i)\right)^2\right]. \tag{10}$$

An average approximation error (in %) was calculated using the following formula: $\bar{A} = n^{-1} \sum |(y_i - \hat{y}_i)/y_i| \cdot 100\%$ for both approximations. For the above formulas, we considered $x = D$, $y = h/D$, and $n = N_{set}$.

We also considered sets of points ($D$, $h/D$) taken together from sets corresponding to craters from Mazrouei et al. (2019) and Losiak et al. (2015). Such sets are denoted in the tables as "Losiak + Mazrouei". In these papers, the authors sometimes used different values of diameters for the same craters. Therefore, the values ($D$, $h/D$) can be a little different for the same crater for sets in Tables 2 and 4 corresponding to different papers. For lines marked by * and located at the bottom of the tables, we consider one point ($D$, $h/D$) for the crater which was considered in both papers. The value of $D$ for such points was taken from Mazrouei et al. (2019). For upper lines marked by "Losiak + Mazrouei" (without *) in Tables 2 and 4 for the same crater, we took two points ($D$, $h/D$), based on the values of diameters of craters presented in both papers. In Tables 3 and 5 we used our estimates of diameters of craters. For several lines of Tables 3 and 5 (for diameters calculated by us), the functions of approximation are plotted in Figs. 5 and 6. The lines in the tables for which there are plots in the figures are marked by = and bold text in the tables.

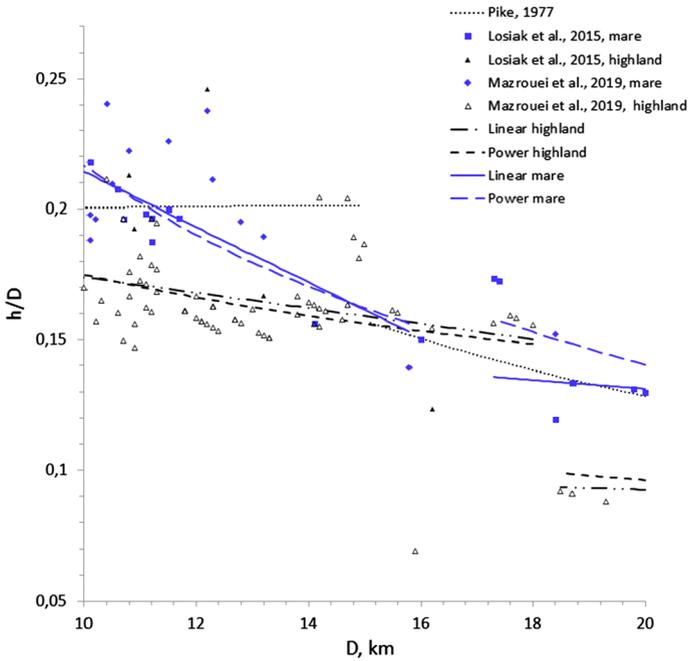

**Fig. 5** Relationship of the *h/D* ratio of a crater depth to its diameter with the diameter for Copernican craters with a diameter $D \leq 19$ km for the highlands and the maria. The Copernican craters are selected based on the data from Mazrouei et al. (2019) and Losiak et al. (2015). Depths of the craters are derived from LOLA LRO altitude profiles. The linear and power approximations for considered Copernican craters are plotted for the mare and highland craters. The formulas used for the approximation are presented in the bold lines of Tables 3 and 5 marked by =. The fit line corresponding to the formula (1), obtained by Pike (1977) for simple craters, is presented for comparison

In Figs. 5 and 6 we present the values of the ratio *h/D* of the depth *h* to the diameter *D* of a crater for $D \leq 19$ km and $D \geq 16$ km, respectively. Points with $16 \leq D \leq 19$ km are on both figures. Considered Copernican craters and their diameters were selected based on the data from Losiak et al. (2015) and Mazrouei et al. (2019). Diameters and depths of the craters were calculated by us based on the LOLA LRO altitude profiles.

## 4.2 The Ratio of the Depth of a Crater to its Diameter at Different Diameters

Depending on the values of *h/D* we sorted the craters as "simple" and "complex" craters. As mentioned in the Introduction, craters are divided into simple and complex according to their morphology, with the addition of intermediate craters. At this point we do not analyze the morphology of craters. Here we separate craters into two groups based only on the ratio *h/D*. From Figs. 5 and 6 "simple" craters mainly have $D < 15$ km and "complex" craters $D > 19$ km, and there could be craters of both types at $15 < D < 19$ km.

The values of *h/D* for some craters with *D* belonging to the transition *D* interval between simple and complex craters can differ noticeably from the values of *h/D* for other craters with close values of *D*. Considerable difference in the values of *h/D* is found also for values of *D* outside the transition interval. Therefore, for some fit lines in Tables 2, 3, 4 and 5, we consider sets of craters with the exclusion of one or two craters for which the value of *h/D*

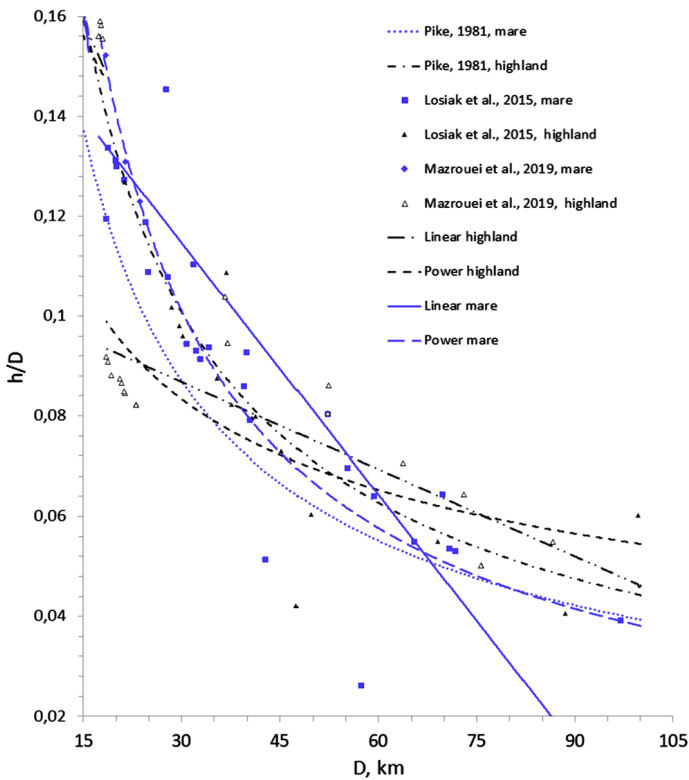

**Fig. 6** Relationship of the *h/D* ratio of a crater depth to its diameter with the diameter *D* for Copernican craters with a diameter $D \geq 20$ km for the highlands and the maria. The selected Copernican craters are from Mazrouei et al. (2019) and Losiak et al. (2015). Depths and diameters of the craters were calculated using LOLA LRO altitude profiles. The linear and power approximations for the mare and highland Copernican craters are plotted. The formulas used for the approximation are presented in the bold lines of Tables 3 and 5 marked by =. The plots corresponding to the formulas (3) and (4), obtained by Pike (1981) for mare and highland complex craters, are presented for comparison

differs noticeably from other values of *h/D* at close values of *D*. For example, in Table 2 for the fit lines (Mazrouei, highland, 10–18) the point ($D=16$, $h/D=0.069$) is present at $N_{set}=69$ and is excluded at $N_{set}=68$. None of the points in sets for the tables are excluded if there are no similar fit lines with different values of $N_{set}$. In Table 2 for the set for the fit line (*Losiak+Mazrouei, highland, 10–18) at $N_{set}=69$, two points ($D=16$, $h/D=0.069$; $D=16$, $h/D=0.125$) are excluded compared to the line at $N_{set}=71$.

For all highland Copernican craters considered in Losiak et al. (2015), we have $h/D \geq 0.16$ at $D \leq 15$ km, $h/D=0.125$ at $D=16$ km, and $h/D \leq 0.122$ at $D \geq 22.1$ km (there are no craters with $16 < D < 22$ km). For all highland Copernican craters considered in Mazrouei et al. (2019), we have $h/D \geq 0.15$ at $D \leq 18$ km if we exclude the point ($D=16$, $h/D=0.069$), and $h/D \leq 0.103$ at $D \geq 19$ km. We can conclude that complex highland Copernican craters have diameters $D \geq 19$ km, though both "simple" and "complex" craters can be at $16 \leq D \leq 18$ km.

In Table 4 the sets for lines (*Losiak+Mazrouei, mare, 19.5–97) with $N_{set}=27$ and $N_{set}=26$ and the sets for lines (*Losiak+Mazrouei, mare, 10–97) with $N_{set}=51$ and $N_{set} = 50$ are differed by the point ($D = 57.3$, $h/D = 0.026$). The same point was excluded in Table 5 for lines "Losiak + Mazrouei" at the range of diameters (19.8–96.9) and 26 craters, and at the range of diameters (17.3–96.9) and 32 craters. The smaller depth of this crater (Taruntius, coordinates: 5.5 N, 46.5 E) can be caused by lava.

Data for the fit lines for highland craters presented in Tables 2 and 3 show that the error of approximation is less than 0.1 at $10 \leq D \leq 18$ km and usually do not exceed 0.15 at $D > 18$ km. Based also on Fig. 5 we assume the separation of highland craters into two groups according to their dependences of $h/D$ on $D$: at $D < 18$ km there are mainly "simple" craters, but some "complex" craters can have diameters $D \geq 16$ km (e.g. the crater with $D = 16$ and $h/D = 0.069$).

For mare craters with $D \leq 14$ km, we have $h/D > 0.15$. Similar to Pike (1981) we presuppose that mare craters with $D < 15$ km to be simple craters. Mare craters with $15 < D < 18$ km are in an agreement with approximation curves for both simple and complex craters. Mare craters with $D > 18$ km are in a better agreement with the curve for complex craters.

In Fig. 1 in Pike (1981) the transition from simple to complex craters occur at $D = 15$ km for the maria, $D = 21$ km for the highlands, and at $15 < D < 21$ km simple highland craters are deeper than complex mare craters of the same diameter. Note that the age of some of the craters considered by Pike (1981) may be older than the age of the Copernicus craters considered by us. Our estimates of the transition diameters between simple and complex craters differ a little from Pike (1981). For example, we consider complex highland craters with $D \geq 19$ km, not with $D \geq 21$ km, as Pike (1981).

Similar to formula (9) let us consider $h/D = k \cdot D^{\alpha}$, where $k$ and $\alpha$ are the coefficients calculated using formulas (10). For simple craters, Pike (1977) obtained $k = 0.196$ and $\alpha = -0.01$. Below in this paragraph, we deal only with the lines corresponding to simple craters. In Tables 3 and 5 for lines "=Losiak + Mazrouei, 10-", we have $k = 0.334$ and $\alpha = -0.281$ for highland craters and $k = 1.124$ and $\alpha = -0.715$ for mare craters. As it is seen from Fig. 5, these power approximations are close to our linear approximations. These approximation lines, especially for mare craters, are at a steeper angle than the line obtained by Pike (1977). For lines "*Losiak + Mazrouei" for craters taken from both Mazrouei et al. (2019) and Losiak et al. (2015), we consider the same sizes of craters as for "Mazrouei" lines in Tables 2 and 4, and the number of simple craters from Mazrouei et al. (2019) is greater than that from Losiak et al. (2015). Therefore, the approximation coefficients for "*Losiak + Mazrouei" sets are close to those for "Mazrouei" sets. For "Losiak" tables' lines, $k > 1$ and the value of $|\alpha|$ is greater than the values of $|\alpha|$ for the lines "*Losiak + Mazrouei" and Pike (1977).

We also used the linear approximation in the form $h/D = b \cdot D + c$, where $b$ and $c$ are coefficients calculated with the use of formulas (8). For simple craters for "=Losiak + Mazrouei, 10-..." sets in Table 5, we have $b = -0.00298$ and $c = 0.2037$ for highland craters, and $b = -0.01056$ and $c = 0.3198$ for mare craters. From these values of $c$ and approximation linear lines in Fig. 5, it is seen that the line for mare craters is higher than the line for highland craters at $D < 15$ km. That is, for the same diameter, such mare simple craters are a little deeper than highland simple craters. The values of $h/D$ for our approximation curves are mainly smaller than the values of the curve by Pike (1977) at $D < 15$ km. Only for mare craters at $D < 11$ km, our line is a little higher than the curve by Pike (1977).

For complex craters (considered below in this paragraph), Pike (1981) obtained $k = 1.028$ and $\alpha = -0.683$ for highland craters and $k = 0.819$ and $\alpha = -0.659$ for mare craters. Kalynn et al. (2013) obtained $k = 1.558$ and $\alpha = -0.746$ for highland craters and $k = 0.870$ and $\alpha = -0.648$ for mare craters. The difference in $k$ and $\alpha$ between the values obtained by Pike and by Kalynn was greater for highland craters than for mare craters. In Tables 3 and 5 for lines "=Losiak+Mazrouei" at $D>17$ km, we have $k=0.277$ and $\alpha=-0.353$ for highland craters, and $k=1.589$ and $\alpha=-0.810$ for mare craters. For highland craters, the values of $k$ and $|\alpha|$ are smaller ($k$ by up to a factor of several) than those for the dependencies obtained by Pike (1981) and Kalynn et al. (2013). For mare craters such values are a little higher than those from these papers. The difference in our lines for highland craters and mare craters can be partly caused by that in Fig. 6 for highland craters there is a group of points with $h/D$ about 0.08–0.09 and $D$ between 18 and 23 km. All mare craters for such $D$ are deeper, with $h/D \geq 0.108$ at $D \leq 28$ km. The above conclusions reached based on Tables 3 and 5 (with our estimates of diameters) can be also made based on Tables 2 and 4, which use the values of diameters taken from Losiak et al. (2015) and Mazrouei et al. (2019).

### 4.3 Comparison of the Depths of Mare and Highland Craters

Our values of $k$ in power approximation were smaller (and $|\alpha|$ was greater) for highland craters than those for mare craters. Our power approximation formulas can be presented in the form similar to formulas (1)–(6):

$$h = 0.277 D^{0.647}, \quad h/D = 0.277 D^{-0.353} \quad \text{for highland craters at} \quad 18.5 \leq D \leq 99.6 \text{ km} \tag{11}$$

and

$$h = 1.589 D^{0.190}, \quad h/D = 1.589 D^{-0.810} \quad \text{for mare craters.} \quad \text{at} \quad 17.3 \leq D \leq 96.9 \text{ km} \tag{12}$$

The linear approximation for the sets "=Losiak+Mazrouei" for complex craters at the above diameters gives $h/D = -0.000580 D + 0.1042$ for highland craters and $h/D = -0.001679 D + 0.1650$ for mare craters. The fit line for mare craters has a greater inclination than the line for highland craters. It is seen from Fig. 6 that for such linear approximation the values of $h/D$ are greater for mare craters than for highland craters at $D<55$ km and are smaller at $D>55$ km. For power approximations for the same sets of craters, the values of $h/D$ for mare craters are greater than for highland craters at $D<46$ km and are smaller at $D>46$ km. Consequently, based on the above fit lines, we can conclude that the value of the diameter for which $h/D$ is greater for mare craters than for highland craters is probably between 45 and 55 km.

We also compared two lines of Tables 2 and 4: (*Losiak+Mazrouei, highland, 16–98.2, $N_{set}=41$) and (*Losiak+Mazrouei, mare, 17–97, $N_{set}=32$). For linear approximations, we have $h/D = -0.00105 D + 0.1335$ for the highland craters and $h/D = -0.00159 D + 0.1598$ for the mare craters. The lines cross each other at $D=48.7$ km. The power approximation $h/D = 0.491 D^{-0.497}$ is for the highland craters, and $h/D = 1.463 D^{-0.789}$ is for mare craters, and these lines cross each other at $D=42$ km. For most lines in Tables 2, 3, 4 and 5 (especially for mare craters), the errors for power approximation were less than those for linear approximation.

Based on analysis of data in Table 1 considered in the previous section, we concluded that this value is about 30 or 40 km. In our opinion, the estimate based on the data from Table 1 can be better than that based on the points of intersections of the above fit lines.

Summarizing these results (also those for simple craters), we conclude that at the same diameter mare craters are deeper than highland craters at a diameter smaller than 30–40 km. For greater diameters, highland craters are deeper. At different diameters, we reached different conclusions on the relative depths of craters on the highlands and the maria. Such difference can be due not only to various properties of the underlying surface, as it was suggested by Kalynn et al. (2013), but also to some differences in the mechanism of formation of craters on the highlands and the maria. The explanation of the differences in depths of Copernican craters on the highlands and the maria at different diameters of the craters can be an item of future studies, especially for those who simulate collisions of impactors with the lunar surface.

From Fig. 4 we see that for formulas obtained by Pike (1981) and Kalynn et al. (2013) the ratio $h/D$ is greater for highland craters than for mare craters for any complex craters. For our estimates, this conclusion is true only for craters with a diameter greater than 30 or 40 km. If we compare the power approximations obtained with the use of formulas (11)-(12) with Pike's formulas (3)-(4), we conclude that for our power approximations of the dependence of $h/D$ on $D$ obtained for complex craters, the values of $h/D$ are greater than those obtained by Pike (1981) at $D > 53$ km for highland craters, and at $D < 80$ km for mare craters. As Pike (1981) considered older craters, then our greater estimates of $h/D$ at $D < 80$ km for mare craters can be due to a faster rate of destruction of such craters with time.

In Fig. 6 the plot shows five points corresponding to highland complex craters close to the line (not shown in the figure) between the points (18.5, 0.092) and (23.1, 0.082). On the continuation of this line, there are three points: (43, 0.052, mare), (47, 0.042, highland) and (57.3, 0.026, mare), which are located far from other points with greater values of $h/D$ at close values of $D$. Maybe the discrepancies in $h/D$ are caused by some differences in layering in the target terrain for this group of craters.

## 5 Conclusions

The relationship between the depth of a crater and its diameter for lunar craters on both the maria and the highlands was compared. The focus was on craters belonging to the Copernican period (i.e., with an estimated age of less than 1.1 billion years). The data on ages of the Copernican craters were taken from the papers by Mazrouei et al. (2019) and Losiak et al. (2015). The depths $h$ and diameters $D$ of the craters were derived from the altitude profiles constructed with the use of the altimeter data obtained by LOLA (the Lunar Orbiter Laser Altimeter), which was a part of the Lunar Reconnaissance Orbiter (LRO) scientific payload. The obtained ratios of crater depths to crater diameters can differ by up to a factor of two for craters with almost the same diameters.

The ratio of the depth of a crater to its diameter ($h/D$) for craters from Mazrouei et al. (2019) and Losiak et al. (2015) is found to differ from those calculated by Pike (1977, 1981). In most cases, the values of the ratio are located below the fit curve obtained in Kalynn et al. (2013) for craters on the highlands, i.e., they are more consistent with data from Pike (1977, 1981) than with data from Kalynn et al. (2013).

The linear and power approximations of $h/D$ on $D$ were made for simple and complex Copernican craters selected from the data in Mazrouei et al. (2019) and Losiak et al. (2015).

For the separation of highland craters into two groups according to their dependences of $h/D$ on $D$, at $D < 18$ km there are mainly "simple" craters, but some "complex" craters can have diameters $D \geq 16$ km.

Mare craters with $D \leq 14$ km have $h > 0.15D$. We agree with Pike (1981) that mare craters with $D < 15$ km are simple craters. Mare craters with $15 < D < 18$ km are in agreement with both approximation curves for simple and complex craters. Depths of mare craters with $D \geq 18$ km are in a better agreement with the approximation curve of $h/D$ versus $D$ for complex craters than for simple craters.

At the same diameter, mare craters are deeper than highland craters at a diameter smaller than 30–40 km. For greater diameters, highland craters are deeper.

The values of $h/D$ for our approximation curves are mainly smaller than the values of the curve by Pike (1977) at $D < 15$ km. Only for mare craters at $D < 11$ km our fit line is a little higher than the curve by Pike (1977).

For our power approximations, the values of $h/D$ obtained for complex craters are greater than those obtained by Pike (1981) at $D > 53$ km for highland craters, and at $D < 80$ km for mare craters.

The conclusions of the paper are the same whether we use our estimates of diameters of craters or we use the diameters presented by other authors.

**Acknowledgements** The authors would like to express their gratitude to R. Aileen Yingst and to an anonymous reviewer for very helpful comments that have considerably improved the manuscript. The work was carried out as a part of the state assignments of the P.K. Sternberg Astronomical Institute of MSU № AAAA-A20-120012990068-0 and the V.I. Vernadsky Institute of Geochemistry and Analytical Chemistry of RAS No. 0137-2020-0004. For S.I., studies of the formulas for the relationship of the ratio of the depth of a crater to its diameter with the diameter were supported by Ministry of Science and Higher Education of the Russian Federation under the Grant 075-15-2020-780 (N13.1902.21.0039).

## Compliance with Ethical Standards

**Conflict of interest** The authors declare that they have no conflict of interest.

## Authors and Affiliations

### E. A. Feoktistova[1]  · S. I. Ipatov[2]

S. I. Ipatov
siipatov@hotmail.com

[1] P.K. Sternberg Astronomical Institute of M.V. Lomonosov Moscow State University, Moscow, Russia

[2] V.I. Vernadsky Institute of Geochemistry and Analytical Chemistry of Russian Academy of Sciences, Moscow, Russia